\documentclass[journal=jacsat,
manuscript=letter,
layout=twocolumn]{achemso}

\usepackage{graphicx}% Include figure files
\usepackage{dcolumn}% Align table columns on decimal point
\usepackage{bm}% bold math
\usepackage{physics}
\usepackage{xcolor}
\usepackage{booktabs}

\usepackage[english]{babel}
\usepackage[%
  colorlinks=true,
  urlcolor=blue,
  linkcolor=blue,
  citecolor=blue
]{hyperref}
\title{  Emergence of Spinmerism for Molecular Spin-Qubits Generation  }% Force line breaks with \\
 \author{Pablo Roseiro} 
 
\author{Louis Petit}

 \author{Vincent Robert}
 \email{vrobert@unistra.fr}
 
 \author{Saad Yalouz}
 \email{yalouzsaad@gmail.com} 
 
 \affiliation{Laboratoire de Chimie Quantique,\\ Institut de Chimie,
CNRS/Université de Strasbourg,\\ 4 rue Blaise Pascal, 67000 Strasbourg, France}%Lines break automatically or can be forced with \\

\begin{document}

\begin{abstract}  
Molecular platforms  are regarded as promising candidates in the generation of units of information for quantum computing. 
Herein, a strategy combining spin-crossover metal ions and radical ligands is proposed from a model Hamiltonian first restricted to exchange interactions. 
Unusual spin states structures emerge from
the linkage of a singlet/triplet commutable metal centre 
with two doublet-radical ligands.  
The ground state nature is modulated by charge transfers and can exhibit a mixture of  
triplet and singlet local metal spin states. 
Besides, the superposition reaches a maximum for $2K_M = K_1 + K_2$, suggesting a necessary competition between the intramolecular $K_M$ and inter-metal-ligand  $K_1$ and $K_2$ direct exchange interactions. 
The results promote \textit{spinmerism}, an original manifestation of quantum entanglement between the spin states of a metal centre and radical ligands. The study provides insights into spin-coupled compounds and inspiration for the development of molecular spin-qubits.

\end{abstract}

\maketitle

\section{Introduction}

% The synthesis and characterization of molecule-based magnetic systems remains an intense research area with applications outcomes such as molecular switches, thermal sensors and long-dreamt information storage devices. Natural building blocks are paramagnetic metal ions which can be coupled through polarizable ligands. Organic radicals have considerably diversified the possibilities offered by transition metal ion complexes, not to mention their ability to bind and stabilise high oxidation states of metal ions~\cite{ghosh2003noninnocence,tomson2011redox,berry2006octahedral,rota2008inspection}. 
%  In self-assemblies, the flexibility inherently attributed to the contacts is likely to modulate the inter-unit interactions. This modulation calls for various schemes of rationalisation ranging from exchange interactions coupling to spin-crossover phenomenon. Evidently, a prerequisite is the presence of a spin-switchable centre, prototypes being spin-crossover ions such as iron(II) or cobalt(II) ($d^6$ and $d^7$, respectively). 

 The synthesis and characterization of molecule-based magnetic systems remains an intense research area with applications outcomes such as molecular switches, thermal sensors and long-dreamt information storage devices for quantum technologies.  The motivations for using such complexes   originate from their apparent long coherence time~\cite{
atzori2016quantum, atzori2016room, bader2014room, graham2014influence, atzori2018structural} and 
% \textcolor{red}{ 
their efficient addressing potential
% }
~\cite{bayliss2020optically,carretta2021perspective,li2021manipulation,nelson2020cnot,thiele2014electrically,hussain2018coherent}. In this context, various types of molecular systems have been investigated ranging from transition metal complexes to organic magnetic molecules~\cite{chiesa2021embedded,maylander2021exploring,nelson2020cnot,cimatti2019vanadyl,zadrozny2015millisecond,gaita2019molecular,lehmann2009quantum,timco2009engineering} to cite but a few.

Natural building blocks are paramagnetic metal ions which can be coupled through polarizable ligands. Organic radicals have considerably diversified the possibilities offered by 3d ions, 
not to mention their ability to bind and stabilise high oxidation states of metal ions~\cite{ghosh2003noninnocence,tomson2011redox,berry2006octahedral,rota2008inspection}. 
 In self-assemblies, the flexibility inherently attributed to the contacts is likely to modulate the inter-unit interactions. This modulation calls for various schemes of rationalisation, ranging from exchange interactions coupling to spin-crossover phenomenon. Evidently, a prerequisite is the presence of spin-switchable units, prototypes being spin-crossover ions such as iron(II) or cobalt(II) ($3d^6$ and $3d^7$, respectively). 
 Similar observations were reported in Prussian blue analogues where the mobility of the counter cation displacement triggers the low-spin Co(III) to high-spin Co(II) transition within the material~\cite{bordage2020influence,cafun2010photomagnetic}

\begin{figure}
    \centering
    \includegraphics[width=8cm]{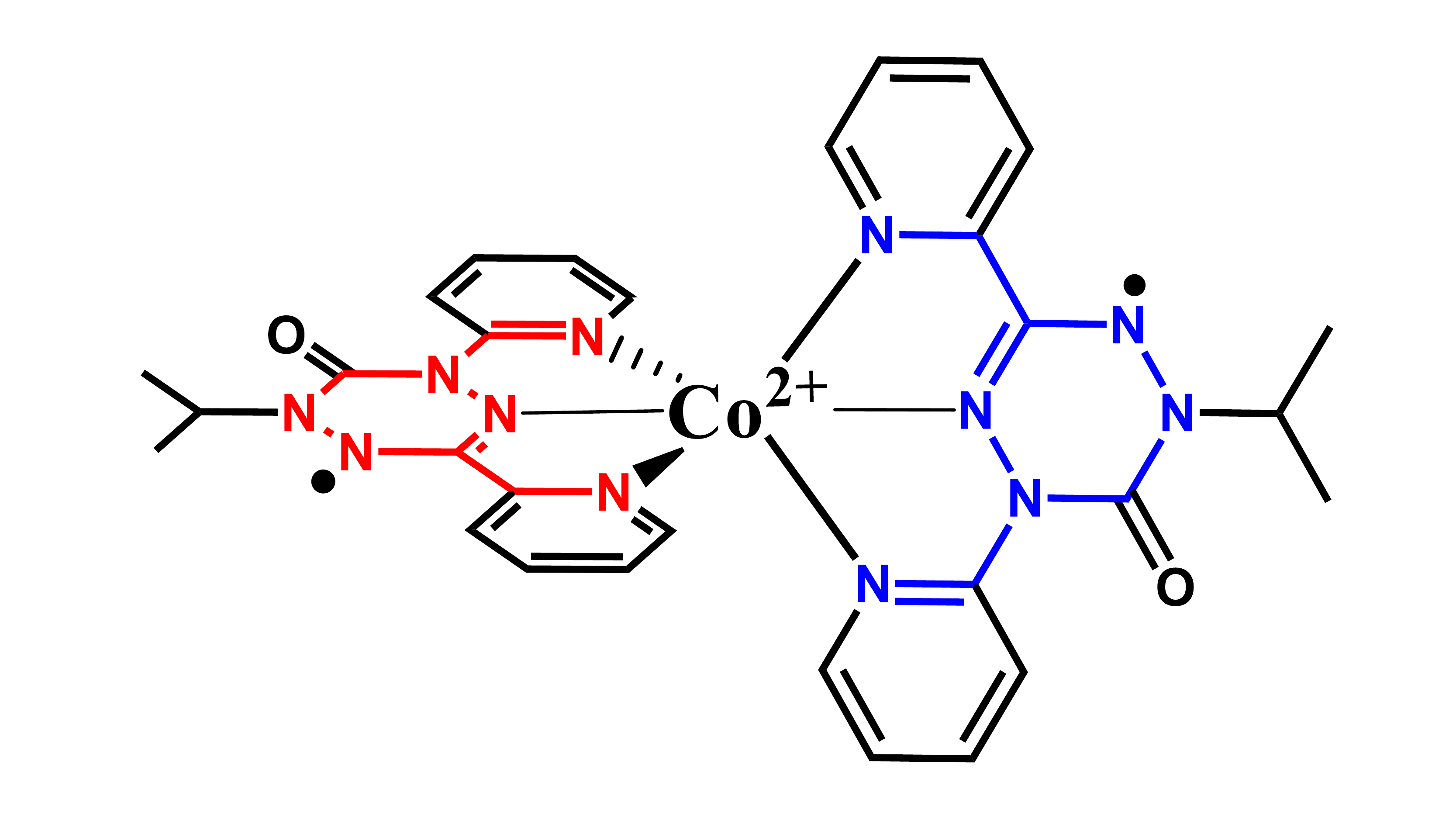}
    \caption{ [Co(dipyvd)$_{2}$]$^{2+}$ (dipyvd = 1-isopropyl-3,5-dipyridyl-6-oxoverdazyl), a representative of the $ML_1L_2$-complexes family. $M$ is a 3d spin-crossover ion, $L_1$ and $L_2$ are radical ligands.    }
    \label{fig:molecule_intro}
\end{figure}

At the crossroad of exchange coupled and spin-crossover compounds, intriguing cobalt(II)-based systems have questioned the traditional pictures emerging from a metal ion, either high-spin or low-spin, in the electrostatic field of neighbouring ligands~\cite{fleming2020valence,roseiro2022combining}. 
The amplitudes of the charge transfers (LMCT, ligand-to-metal charge transfers, and MLCT, metal-to-ligand charge transfers) determine the geometry, spectroscopy and the spin states orderings in such coordination complexes.
Since the ligand field includes Coulomb and exchange contributions in a complex built on spin-coupled partners, one may wonder whether different local spin states may coexist on the metal ion.
The introduction of radical ligands may indeed disrupt the assumption of a given spin state on the metal centre.  Recently, \textit{ab initio} calculations~\cite{roseiro2022combining} have supported such speculation in 
a Cobalt-verdazyl coordination compound~\cite{fleming2020valence} (see Figure~\ref{fig:molecule_intro}) where coherent explanation had remained elusive so far.

Despite its robustness, deviations from the Heisenberg-Dirac-Van Vleck spin Hamiltonian were reported and theoretically explained by the appearance of the so-called non-Hund forms. These contributions were first reported in the study of manganese oxides~\cite{bastardis2007microscopic} and later evoked to account for non-Heisenberg behaviours\cite{anderson1955considerations,bastardis2008isotropic,bastardis2007microscopic,papaefthymiou1987moessbauer}. The importance of the three-body operator in three-centre systems was stressed as a major source of deviation. Nevertheless, the direct exchange contributions in these systems were considered as negligible, whereas ferromagnetic interactions are observed in verdazyl-based inorganic compounds.~\cite{fleming2020valence} %Therefore,  a description dominated by direct exchange couplings may dominate,  and include in a second step the super-exchange contributions following a perturbation scheme.
Therefore, direct exchange couplings may dominate, and super-exchange contributions should then be included in a second step. 

Prompted by the originality of coordination compounds built on spin-crossover ions and radical ligands, we question the use of such complexes for the development of new quantum unit of information, \textit{i.e.} qubits (for quantum of bits). 
To this purpose, we derive a model Hamiltonian to account for the recently suggested \textit{spinmerism} effect~\cite{roseiro2022combining}  
and to motivate its potential use for qubit implementation. 
If practically accessible, the tunability of the metal local spin states (via the \textit{spinmerism} phenomenon\cite{roseiro2022combining}) could provide an innovative way to encode and manipulate information for molecule-based quantum computers. 

\begin{figure*}
    \centering
    \includegraphics[width=16cm]{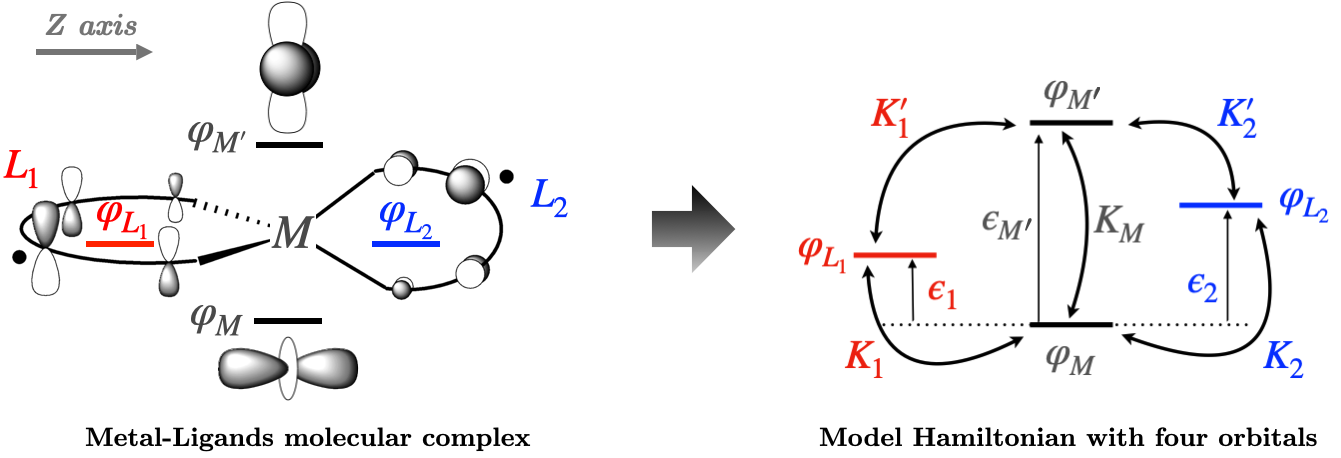}
    \caption{ \textbf{Illustration of a $C_{2v}$-like symmetry $ML_1L_2$ model system}. \textbf{Left panel :} representation of the molecular complex composed of a central spin-crossover ion $M$ ($S_M = 0$ or $1$) coordinated to two radical ligands $L_1$ and $L_2$ ($S_{L_1} = S_{L_2} = 1/2$).\textbf{ Right panel :} Associated model Hamiltonian composed of four orbitals. The orbital labels, orbital energies and non-zero direct exchange integrals involved in the zeroth-order Hamiltonian are shown. 
    %The indexes refer to the ligands $L_1$ and $L_2$, holding orbitals $\varphi_{L_1}$ and $\varphi_{L_2}$, respectively. 
    % The metal centre direct exchange $K_M$ value is used as a reference.
    }
    \label{fig:Model_illustration}
\end{figure*}

This view concentrates the effort on extracting some rules 
and synthetic strategy
following a zeroth-order description based on direct exchange interactions. Therefore, a three-site model system is considered, including a spin-versatile metal ion $S_M = 0$ or $1$ (\textit{e.g.} Ni$^{2+}$ ion in an octahedral environment) and two radical ligands $S_{L_1} = S_{L_2} = 1/2$ (see Figure~\ref{fig:Model_illustration}). The eigenfunctions of the model Hamiltonian $\hat{H}_0$ written on the neutral configurations (\textit{i.e.} singly occupied on-site orbitals) are decomposed on the local spin states. The contributions of the $S_M = 0$ and $S_M = 1$ components are evaluated in the ground and excited states as a function of the exchange interactions. A key parameter is the metal exchange interaction that not only governs the positions of the non-Hund forms, but also elementary rules that are derived.  Then, the energies are corrected using second-order perturbation theory to include charge transfers. These contributions account for the fluctuations that must be introduced to go beyond a mean-field picture.  

The originality of this work stems from the combination of a spin-crossover ion and organic radical ligands where the weights of the metal local spin states can be modulated. 
 The use of molecular-spin degrees of freedom to encode and/or manipulate quantum information onto magnetic molecules remains a growing field of research.

\section{Theory } 

\subsection{Description of the model} 

The system consists of four electrons distributed on ligand- and metal-centred orbitals referred to as $\varphi_{L_1}$, $\varphi_{L_2}$, $\varphi_{M}$ and $\varphi_{M'}$ (see Figure~\ref{fig:Model_illustration}). In reference to coordination chemistry compounds, the $\varphi_{M}$ and $\varphi_{M'}$ orbitals correspond to the $d_{z^2}$ and $d_{x^2-y^2}$ singly occupied atomic orbitals of a tetrahedral $d^2$ (or octahedral-like $d^8$) ion. The orbitals $\varphi_{L_1}$ and $\varphi_{L_2}$ may be seen as the $\pi$-frontier molecular orbitals localised on the radical ligands.
In the total spin projection $M_S = 0$ manifold, the zeroth-order Hamiltonian $\hat{H}_0 = \hat{P}_{} \hat{H} \hat{P}$ is built from the full Hamiltonian $\hat{H}$ and  the projector $\hat{P} = \sum_{\alpha }  \ketbra{\Phi_\alpha}{\Phi_{\alpha}} $ over the subset of six neutral configurations $\lbrace | \Phi_\alpha \rangle \rbrace $ defined as
\begin{equation}
\begin{split}
    \lbrace \ket{\Phi_\alpha} \rbrace  =  
    \bigg\lbrace 
    &|\overline{\varphi}_{L_1}\varphi_{M}\varphi_{M'} \overline{\varphi}_{L_2}|, 
    |\varphi_{L_1} \varphi_{M} \overline{\varphi}_{M'} \overline{\varphi}_{L_2}|,\\
    &|\overline{\varphi}_{L_1}\varphi_{M}\overline{\varphi}_{M'} {\varphi}_{L_2}|, 
    |\varphi_{L_1}\overline{\varphi}_{M}\varphi_{M'} \overline{\varphi}_{L_2}|, \\
    &|\overline{\varphi}_{L_1} \overline{\varphi}_{M}\varphi_{M'} \varphi_{L_2}|, 
    |\varphi_{L_1}\overline{\varphi}_{M} \overline{\varphi}_{M'} {\varphi}_{L_2}|  
     \bigg\rbrace .
\end{split}
\end{equation} 
This subspace  will be referred to as the inner (or model) $\alpha$-space (following regular notations as used in Ref.~\cite{lindgren2012atomic}) characterised by singly-occupied orbitals. The resulting zeroth-order Hamiltonian takes the following form 
\begin{equation}
    \hat{H}_{0} = \sum_{\alpha,\alpha'} \textbf{H}_{\alpha\alpha'} \ketbra{\Phi_\alpha}{\Phi_{\alpha'}} 
    \label{eq:H0}
\end{equation} 
 From spin coupling algebra $\hat{S} = \hat{S}_M + \hat{S}_{L_1} + \hat{S}_{L_2}$, two singlet, three triplet and one quintuplet eigenstates are generated.
 %for the $ML_1L_2$ system. 
The zeroth-order Hamiltonian matrix elements ${\bf H}_{\alpha\alpha'}$ introduced in Eq.~\ref{eq:H0} are functions of the on-site energies (one-electron contributions) and positively defined two-electron integrals. The one-electron energies are referenced to the $\varphi_{M}$ orbital energy as $\epsilon_{M'}$, $\epsilon_1$ and $\epsilon_2$ for the $\varphi_{M'}$, $\varphi_{L_1}$ and $\varphi_{L_2}$ orbitals, respectively (see Figure~\ref{fig:Model_illustration}). Evidently, the single-occupation of the orbitals in the  $\lbrace \ket{\Phi_\alpha} \rbrace $ configurations leads to a common $  \epsilon_{M'} + \epsilon_1 + \epsilon_2$  value on the diagonal elements of the six-by-six matrix. 
The off-diagonal matrix elements are linear combinations of the two-electron integrals. 
% expressed in the spin-orbitals basis  as :
% \begin{equation}
%     (\varphi_p \varphi_q,\varphi_r \varphi_s) = \iint dx_1dx_2 \frac{\varphi_p^*(x_1)  \varphi_q(x_1)  \varphi_r^*(x_2)  \varphi_s(x_2)}{r_{12}}
% \end{equation}  
% where the integration is carried out on spatial and spin coordinates. 
The system may equivalently be examined from two subunits, namely the metal ion centre $M$ and the ligands pair $L_1L_2$. For the former, the energy difference between the Hund triplet and non-Hund singlet states is $2K_M$, where $K_M 
% = ( \varphi_{M}\varphi_{M'}, \varphi_{M'}\varphi_{M})
$ is the atomic exchange interaction. This is a dominant contribution in free ions, but the energy splitting is evidently much affected by the field generated by the ligands. In spin-crossover compounds, the low-spin and high-spin states lie close enough in energy to observe a transition for moderate ligands field modification. In contrast, one would anticipate a negligible Ligand-Ligand  
% $(\varphi_{L_1}\varphi_{L_2},\varphi_{L_2}\varphi_{L_1})$
exchange integral in synthetic compounds with $L_1$ and $L_2$ in trans position (see Figure~\ref{fig:Model_illustration}). Thus, this integral was set to zero in our model.

After diagonalizing $\hat{H}_0$,
% (given in Eq.~\ref{eq:H0}) 
% that sets the $ \ket{S = 2, M_S = 0}$ quintuplet state as the ground state (Hund’s rule),
the associated eigenvectors $\ket{\Psi}$ (with unperturbed energy $E_\Psi$) were projected onto the local singlet and triplet states of the $M$ and $L_1L_2$ subunits. The procedure uses the standard Clebsch-Gordan coefficients algebra~\cite{kahn1993molecular}. This transformation allows one to evaluate the singlet and triplet weights in the six different states with respect to the parametrization of the model. In the following, all basis set vectors are written as $\ket{S, S_M, S_L}$ where $S$ is the total spin state. $S_M$ and $S_L$ stand for the local spin values on the metal and the ligands pair, respectively. The control of the amount of $S_M = 0$ or $1$ (and $S_L = 0$ or $1$ on the ligands pair $L_1L_2$) in the $\ket{S, S_M, S_L}$ wavefunctions makes this class of compounds particularly interesting in molecular magnetism and might enrich the panel of molecular spin-qubits candidates.

\subsection{Perturbation Theory}

After evaluating the eigenstates $\ket{\Psi}$ of the unperturbed Hamiltonian $\hat{H}_0$, the associated zeroth-order energies $E_\Psi$ were corrected using second-order perturbation theory to go beyond the mean-field description of the metal-ligands interactions. The fluctuations introducing the electron-electron interactions correspond to charge transfers between the metal centre and the ligands. Following Ref.~\cite{lindgren2012atomic}, we introduced then the so-called outer $\beta$-space as built from the subset of eight LMCT and eight MLCT perturber configurations $\lbrace \ket{\Phi_\beta} \rbrace$. The interaction between the inner $\alpha$-space and outer $\beta$-space was limited to single charge transfers couplings formally modeled by an interaction Hamiltonian $ \hat{V}$ containing a single one-electron hopping integral noted $t$. In addition, this Hamiltonian was extended to incorporate on-site repulsion parameters $U_M$ (for metal)
% =(\varphi_{M'}\varphi_{M'},\varphi_{M'}\varphi_{M'})=(\varphi_M\varphi_M ,\varphi_M\varphi_M)$
and $U_L$ (for each ligand).
% =(\varphi_L\varphi_L,\varphi_L\varphi_L)$
% Note that the quintuplet eigenstate energy of $\hat{H}_0$ is evidently not affected by charge transfers, whereas lower-spin multiplicities are likely to be stabilised.
The energy correction brought by perturbation theory up to second-order reads
\begin{equation}
    E^{PT2}_{\Psi} = E_{\Psi} + \sum_{ \beta}^\text{outer-space}  \frac{|\bra{\Phi_\beta} \hat{V} \ket{\Psi}|^2}{E_{\Psi} - E_\beta }
\end{equation}
where   
$E_\beta = \bra{\Phi_\beta} (\hat{H_0} + \hat{V})\ket{\Phi_\beta}$ is the energy of a given configuration $\ket{\Phi_\beta}$.  
Beyond energy corrections, let us stress that the spin states decomposition is also affected through first-order wavefunction modifications. The perturbers consist of local spin $1/2$ states which modify the projection. Nevertheless, the contracted structure leaves the relative weights in the model $\alpha$-space unchanged.

% \saad{ OK VINCENT ! Je laisse a Pablo le choix de garder ou nan ce qui suit en fonction de s'il souhaite enrichir plus ;-) } 

% \textbf{
% Practically, estimations of the model parameters can be extracted from the PSI4 suite of programs. Open-shell Hartree-Fock calculations were carried out on an isolated Ni$^{2+}$ ion, and a hypothetical Ni(vdz)$_2$ coordination compound. These values were expressed in KM unit, and then varied to foresee different possible regimes exhibiting low-energy states with a spinmerism manifestation.}  

\section{Numerical results}

All our calculations were performed by fixing the $K_M$ value to unity. In the present description, this is the leading parameter that is expected to become vanishingly small for spin-crossover ions. The spin states structure is first analyzed from the zeroth-order Hamiltonian $\hat{H}_0$. Subsequently, the spin states energies are corrected by the outer $\beta$-space perturbers to foresee the low-energy spectroscopy of our model system. For $K_M \gg K_i$, $K'_i$ , the spectroscopy splits into two sub-sets identified by the $S_M$ value, $S_M=1$ and $S_M=0$. This is the standard situation based on the atomic Hund’s states in coordination chemistry compounds. However, the picture might be much different when the direct exchange couplings compete (\textit{i.e.} all $K_i$ and $K'_i$ of the order of $K_M$) and satisfy particular conditions. In the absence of symmetry (\textit{i.e.} $ML_1L_2$ compound), the direct exchange couplings are all different and the diagonalization of $\hat{H}_0$ produces eigenvectors which project on pure $S_M=0$ or $S_M=1$ states on the metal ion. A strict separation between the Hund and non-Hund states is observed. The systems ruled by such Hamiltonian fall in the traditional category of metal ion complexes where the metal spin state $S_M$ is a good quantum number. Nevertheless, this particular picture is deeply modified as soon as a higher symmetry is introduced by reducing the number of parameters.

Let us first examine the spin states structures for $K_1 = K_2$, while maintaining $K'_1 \neq K'_2$. This scenario is expected in spiro-like geometry where the interactions are invariant along the $z$-axis for similar ligands 
$K_1 = K_{\varphi_{L_1},z^2} = K_{\varphi_{L_2},z^2} = K_2$  
%$ \varphi_{L_1},\varphi_{L_1}d_{z^2})=(d_{z^2}\varphi_{L_2}, \varphi_{L_2}d_{z^2} )$ 
%$(d_{z^2}\varphi_{L_1},\varphi_{L_1}d_{z^2})=(d_{z^2}\varphi_{L_2}, \varphi_{L_2}d_{z^2} )$ 
whereas they significantly differ in the perpendicular $xy$-plane 
%$((d_{x^2-y^2}\varphi_{L_1}, \varphi_{L_1}d_{x^2-y^2}) \neq 0)$, $((d_{x^2-y^2}\varphi_{L_2}, \varphi_{L_2}d_{x^2-y^2}) \simeq 0 )$. 
$(K'_1 = K_{\varphi_{L_1},x^2-y^2} \neq 0$ and $K'_2 = K_{\varphi_{L_2},x^2-y^2} \simeq 0 )$.
Both singlet and triplet manifolds exhibit combinations of local singlet and triplet states. In the following, the amplitudes on the $\ket{S=1, S_M, S_L}$ and $\ket{S=0, S_M, S_L}$ configurations are written as $\lambda_{S_MS_L}$ and $\mu_{S_MS_L}$, respectively. Thus, the weight on the high-spin ($S_M=1$) metal centre for the triplet eigenstates given by 
\begin{equation}
\begin{split}
   \ket{S=1, S_M, S_L} =  \lambda_{11} \ket{S=1, S_M=1, S_L=1} 
    \\+ \lambda_{10} \ket{S=1, S_M=1, S_L=0} 
    \\+ \lambda_{01} \ket{S=1, S_M=0, S_L=1}
\end{split}
\label{eq:triplet_state}
\end{equation} 
reads $\lambda_{11}^2+ \lambda_{10}^2$, a value which may differ from zero or one (see Table~\ref{table:1}). One should note that this mixture does not result from spin-orbit coupling effects which are totally absent in the present description. Such a spin structure that incorporates high-spin ($S_M=1$) and low-spin ($S_M=0$) states on the metal centre is expected from spin algebra, but deeply contrasts with the traditional views on inorganic compounds.  In analogy with the mesomerism effect that accounts for electronic charge delocalization, the model %stresses 
highlights the so-called \textit{spinmerism} phenomenon~\cite{roseiro2022combining} that involves the spin degree of freedom. Finally, the appearance of the $\ket{S=1, S_M=1, S_L=0}$ and $\ket{S=1, S_M=0, S_L=1}$ contributions ($\lambda_{10}$ and $\lambda_{01}$ amplitudes, respectively) stresses the arising of entanglement in the state description. Quantum entanglement is here reflected by the correlation arising between both ligands and metal local spin states $S_L$ and $S_M$ that naturally adjust to fulfil a spin $S=1$ for the full molecular system. Note that, from a pure chemist point of view, spin entanglement would represent a rather unusual picture especially in the case of coordination chemistry compounds where it is usually assumed that local metal and ligand spins states are fixed. This hypothesis however conflicts with fundamental spin and angular momentum algebra which in practice does not forbid the arising of such a feature between two interacting spin sub-systems~\cite{breuer2005entanglement,breuer2005state}. 
%In the following, our numerical results will concretely support the existence of spin entanglement in metal-ligand complexes (within the limit of the model system we are considering in this work).
%  In the following, our model will further support the existence of spin entanglement in metal-ligand complexes.

\begin{table}[!ht]  
  \centering
  \begin{tabular}{c|c|c}
    \toprule
          & $K_1=K_2=0.25$ & $K_1=K_2=0.50$  \\ \midrule
    $S=1$ & $81\%$    & $90\%$   \\
    $S=0$ & $19\%$    & $10\%$    \\ \bottomrule
  \end{tabular}
  \caption{Metal triplet and singlet proportions in the second lowest lying triplet state (see Eq.~\ref{eq:triplet_state}),
%   $\lambda_{11}\ket{S=1, S_M=1, S_L=1} + \lambda_{10}\ket{S=1, S_M=1, S_L=0} + \lambda_{01}\ket{S=1, S_M=0, S_L=1}$. 
$K'_1 = 0.60$, $K'_2 = 0.80$
   ($K_M$  unit).}
  \label{table:1}
\end{table}

Moving to Td-symmetry  compounds
%$ML_1L_2$ 
($L_1 = L_2$)
characterised by $K_1 = K_2$ and $K'_1 = K'_2$, the spin states structure gets further modified. Whereas one triplet state simply reads $ \ket{S=1, S_M=1, S_L=0}$, the other two exhibit a systematic combination as 
% $\ket{S=1, S_M=1, S_L=1}$ and $\ket{S=1, S_M=0, S_L=1}$ components,
\begin{equation}
\begin{split}
       \ket{S=1, S_M, S_L} &=  \lambda_{11}\ket{S=1, S_M=1, S_L=1} \\
       &+ \lambda_{01}\ket{S=1, S_M=0, S_L=1}
\end{split}
\label{eq:second_triplet}
\end{equation} 
The spectroscopy incorporates local high-spin ($S_M=1$) and low-spin states ($S_M=0$) on the metal centre whilst the ligand pair remains $S_L=1$. As seen in Table~\ref{table:deux} for $K'_1 = K'_2 = 0.75$, the proportions are much affected by any modification of the $K_1 = K_2$ value. In practice, exchange integrals are very sensitive to the structure (interatomic bond distances) and the chemical details of the radical ligands. Therefore, one may expect to modify the superposition of metal spin states induced by structural deformations on the coordination complex. In practice, this structural modulation of the system would offer a possible way to encode, and to manipulate, information onto the spin-degree of freedom. 
%in coordination chemistry compounds associated with the metal spin state. 
The molecular complex behaves as a molecular spin-qubit carrying a quantum superposition of local spin states on the metal with tunable amplitudes.

\begin{table}[!ht]  
  \centering
  \begin{tabular}{c|c|c}
    \toprule
                   & $K_1=K_2=0.25$ & $K_1=K_2=0.50$  \\ \midrule
    $S=1$ &  $79\%$   & $86\%$   \\
    $S=0$ & $21\%$    & $14\%$    \\ \bottomrule
  \end{tabular}
  \caption{Metal triplet and singlet proportions in the second lowest lying triplet state (see Eq.~\ref{eq:triplet_state}).
%   $\lambda_{11}\ket{S=1, S_M=1, S_L=1} + \lambda_{10}\ket{S=1, S_M=1, S_L=0} + \lambda_{01}\ket{S=1, S_M=0, S_L=1}$.
  $K'_1 = K'_2 = 0.75 $ ($K_M$ unit).}
  \label{table:deux}
\end{table}

\begin{figure}
    \centering
    \includegraphics[width=8cm]{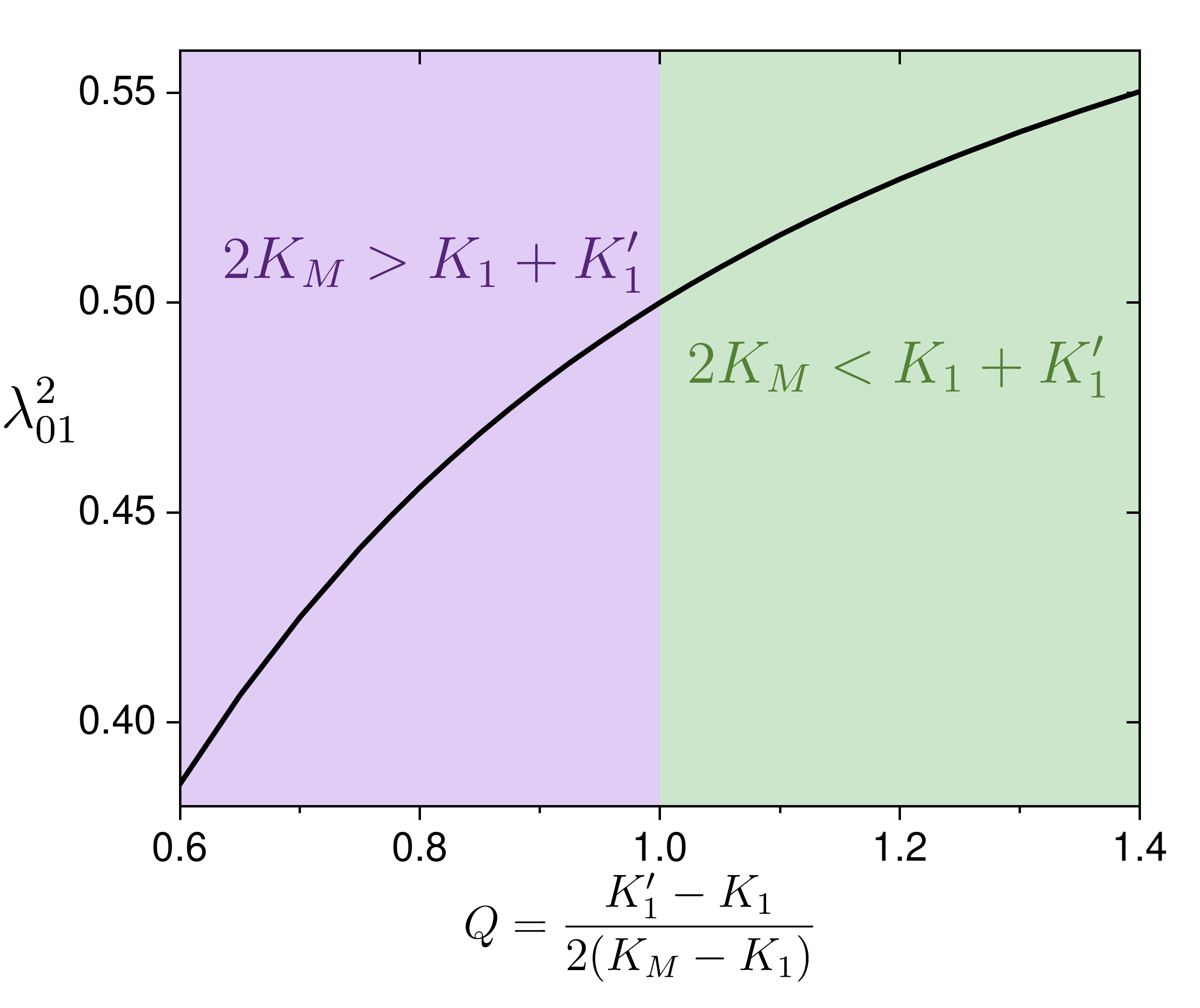}
    \caption{Variations of the $S_M = 0$ weight $\lambda_{01}^2$ as a function of the dimensionless parameter $Q$ (see Eq.~\ref{eq:Q_definition}) in the second lowest lying triplet state (see Eq.~\ref{eq:second_triplet})
    % $\lambda_{11} \ket{S=1, S_M=1, S_L=1} + \lambda_{01}\ket{S=1, S_M=0, S_L=1}$
    for a Td compound characterised by $K_1 = K_2$  and $K'_1 = K'_2$. The mixing is maximal ($\lambda_{11}^2=\lambda_{01}^2=1/2$) for $Q = 1$ (\textit{i.e.}  $2K_M = K_1 + K'_1$).}
    \label{fig:figure_Q}
\end{figure}

Finally, 
%in left-right symmetrical compounds $ML_1L_2$, 
it can be shown that 
the mixing reaches a maximum
$\lambda_{11}^2 = \lambda_{01}^2 $  (equal weights on the $S_M=1$ and $S_M=0$) for 
a first rule $2K_M = K_1 + K'_1$ (see Supporting Information). 
%This condition is difficult to achieve from a synthetic point of view. However, it stresses the prerequisite that at least one ligand-metal direct exchange coupling should be comparable to the intra-atomic exchange value. 
Even though this condition is difficult to achieve from a synthetic point of view, it suggests that at least one ligand-metal direct exchange coupling should be comparable to $K_M$.
%the intra-atomic exchange value.
Any deviation from $2K_M = K_1 + K'_1$ leads to variations of the metal triplet state component $\lambda_{11}^2$   (and singlet $\lambda_{01}^2$) which are shown in Figure~\ref{fig:figure_Q} as a function of the dimensionless parameter
\begin{equation}
    Q      =     \frac{  K'_1  - K_1  }{  2(  K_M - K_1 )  }.
    \label{eq:Q_definition}
\end{equation} 
Along this representation, the associated variations do not depend on the $K_M$ value. For  $Q < 1$, \textit{i.e.} $2K_M > K_1 + K'_1$,  the second lowest lying triplet state is dominated by the $\ket{S=1, S_M=1, S_L=1}$ configuration, whereas  $\ket{S=1, S_M=0, S_L=1}$ is the leading one for $Q > 1$. In the vicinity of $Q = 0.7$,  the changes reach up to $3.5\%$ for deviations smaller than $10\%$. Therefore, any geometrical change induced by external stimuli (\textit{e.g.} pressure, temperature) is likely to deeply modify the spin state structure whatever the $K_M$ value. This observation makes this class of compounds particularly appealing in the generation of innovative spin-qubits.

Moving away from the %left-right symmetry, 
Td-symmetry,
we then examined the $K_1 = K_2 = K'_1$ situation. While the mixing occurred in the triplet manifold, the singlet states are now the intriguing ones. As a matter of fact, superpositions
\begin{equation}
\begin{split}
      \ket{S=0,S_M,S_L} &=   \mu_{11} \ket{S=0, S_M=1, S_L=1} \\
                        &+ \mu_{00}\ket{S=0, S_M=0, S_L=0}
\end{split}
\end{equation}
% $\mu_{01} \ket{S=0, S_M=1, S_L=1} + \mu_{00}\ket{S=0, S_M=0, S_L=0}$ 
are observed. A marked difference is the appearance of different spin states on both the metal centre and the ligands pair. The entanglement between these open-shell subunits reaches a maximum $\mu_{11}^2=\mu_{00}^2$   for $2K_M = 3K_2 + K'_2$. Since all exchange values are positive, the condition $K_2 < 2K_M/3$ is necessary for this equality to be fulfilled. 
%However,
For $K_2$ values larger 
%that 
than $2K_M/3$ , the relative weights ratio is reduced until
 $2K_M = 2K_2 + 2K'_2$, a second rule where $3\mu_{11}^2=\mu_{00}^2$ (see Supporting Information). 
%Beyond these particular ratio values which are reflections of the underlying spin algebra, the possibility to address the local spin states superposition makes these molecular coordination compounds interesting targets for qubits generation.
Such condition displayed by less symmetrical $ML_1L_2$ compounds offers another possibility to address the local spin states superposition. 
%makes these molecular coordination compounds interesting targets for qubits generation.

\begin{figure}
    \centering
    \includegraphics[width=9cm]{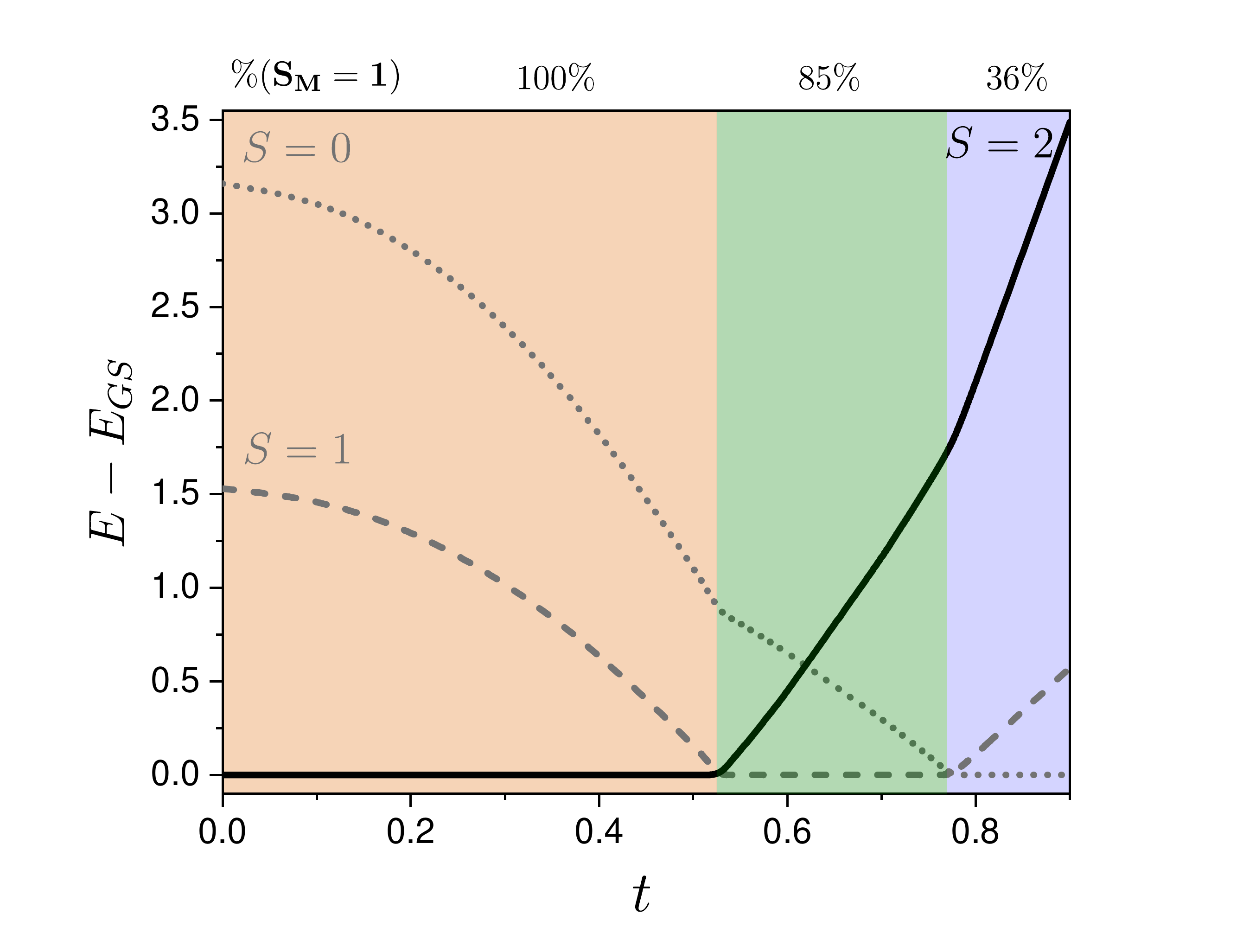}
    \caption{Low-lying states energies of a $ML_1L_2$ system as a function of the hopping parameter $t$ in $K_M$ unit. The $\hat{H}_0$ energies are corrected using second-order perturbation theory with the MLCT and LMCT as perturbers with fixed parameters ($K_M$ unit): $U_L = 2.95$, $U_M = 4.0$, $\epsilon_{M'} =0.80$, $\epsilon_1= -2.80$ and $\epsilon_2=-1.50$, $K_1 = 0.35$,  $K_2 = 0.10$, $K'_1 = 0.58$ and $K'_2 = 0.75$. The ground state energy is used as a reference. }
    \label{fig:perturbation}
\end{figure}

At this stage, the description concentrates on the $\hat{H}_0$ eigenvectors analysis, leaving out the important electronic correlation effects. Therefore, the energies were corrected using second-order perturbation theory accounting for charge fluctuations and
%suggesting 
depicting a more realistic %description of the 
electronic structure.
% Eight LMCT and eight MLCT are identified and contribute to the modulation of the $\hat{H}_0$ eigenvalues.
Such framework is applicable for large enough energy separations between the $\hat{H}_0$ and the perturbers energies, with respect to the hopping integral $t$. Our picture is not valid for systems governed by superexchange contributions (strong field regime) but applicable to intermediate ligand field regimes where several spin multiplicities compete (\textit{i.e.} spin-crossover compounds). The relative weights of the $S_M = 1$ and $S_M = 0$ states are not affected by the outer $\beta$-space, but the spin multiplicity of the ground state is likely to be changed. 
%This is the scenario given in  
As seen in Figure~\ref{fig:perturbation}, the 
ground state switches
from quintet to successively triplet, and singlet as the hopping integral value is increased.
% For $t = 0.52$ ($K_M$ unit), the energy correction is calculated $33\%$  and the triplet ground state is dominated by a $S_M = 1$ metal spin state ($85\%$).
For $t = 0.52$, the energy correction to the $S=1$ state is calculated $33\%$. This triplet becomes the ground state and is dominated by a $S_M = 1$ 
%meta 
spin state ($85\%$).
As $t$ is further increased, the ground state switches to a singlet exhibiting a $36\%$ proportion on the local $S_M = 1$. Let us mention that a perturbative treatment in this regime is more than questionable but the picture survives. Not only is the nature of the ground state sensitive to the strength of the ligand field following traditional pictures, but the metal centre  spin states contributions are significantly modified.

Evidently, any realistic system includes both direct exchange and charge transfers contributions which compete to ultimately dictate the ground state and low-lying excited states. However, our model sets the stage to foresee ground states where the local spin on the metal centre is not uniquely defined, being a superposition of different spin multiplicities. The presence of open-shell ligands as entangled partners in the coordination sphere is a prerequisite for this manifestation. Thus, the variability offered by organic radicals combined with mid-series metal ions should give access to original compounds with fundamental and applied interests.

\section{Conclusion}

The spin states structure of a four electron/four orbital coordination compound $ML_1L_2$ built on a spin-crossover metal ion ($M = d^2$ or $d^8$ ion such as Ni$^{2+}$) and radical ligands ($L_1$ and $L_2$ such as oxoverdazyl) was examined on the basis of a model Hamiltonian. The zeroth-order structure includes the direct exchange interactions, whereas the LMCT and MLCT accounting for superexchange interactions were treated using second-order perturbation theory. From the spin state versatility of the metal ion ($S_M = 0$ or $1$), the spin states of the complex combine different local spin multiplicities. Depending on the relative values of the direct exchange interactions, the 
%eigenvalues 
eigenfunctions 
of the  zeroth-order Hamiltonian can reach an equal mixture of the $S_M = 0$ and $ 1$ metal states entangled with ligands $S_{L} = 1$ and $0$ states. Spin projection gives rise to rules involving the metal atomic exchange interaction $K_M$ and the sum of the ligand-metal $K_1$ and $K_2$. Despite its simplicity, the model stresses that under specific conditions (spin-crossover ion ferromagnetically interacting with radical ligands) superpositions of local spin states are observed and possibly varied. Evidently, such manifestation of entanglement is anticipated from standard spin algebra. However, conditions for superposition of states are suggested here and enlarge the traditional views in coordination chemistry compounds that usually decide on a given spin state. By experimentally probing the local spin density, such molecular compounds might become original targets for spin-qubits generation. 

\bibliography{biblio}%  Produces the bibliography via BibTeX.

\end{document}